\documentclass{aip-cp}

\usepackage[numbers]{natbib}
\usepackage{rotating}
\usepackage{graphicx}
\usepackage{txfonts}
\usepackage{amsfonts,amssymb}
\usepackage{epstopdf,color,comment,pbox}

\def\beq{\begin{equation}}
\def\eeq{\end{equation}}
\def\bit{\begin{itemize}}
\def\eit{\end{itemize}}
\def\ben{\begin{enumerate}}
\def\een{\end{enumerate}}
\def\btab{\begin{tabular}}
\def\etab{\end{tabular}}
\def\btbl{\begin{table}}
\def\etbl{\end{table}}
\def\bfig{\begin{figure}[htb]}
\def\efig{\end{figure}}
\def\bpic{\begin{picture}}
\def\epic{\end{picture}}
\def\bery{\begin{array}}
\def\ery{\end{array}}

\newcommand{\eqref}[1]{(\ref{#1})}
\newcommand{\vev}[1]{\left\langle#1\right\rangle}
\newcommand{\Leq}{\leqslant}

\newcommand{\ML}[1]{\color{green} [ML:#1] \color{black}}

\begin{document}

\title{Light Leptonic New Physics at the Precision Frontier\footnote{To appear in the proceedings of PPC2015, the 9th International Conference on Interconnections between Particle Physics and Cosmology}}

\author[aff1]{Matthias Le Dall\corref{cor1}}

\affil[aff1]{Department of Physics and Astronomy, University of Victoria, Victoria, BC, CANADA.}
\corresp[cor1]{mledall@uvic.ca}

\maketitle
\begin{abstract}
	Precision probes of new physics are often interpreted through their indirect sensitivity to short-distance scales. In this proceedings contribution, we focus on the question of 
which precision observables, at current sensitivity levels, allow for an interpretation via either short-distance new physics or consistent models of long-distance new physics, 
weakly coupled to the Standard Model. The electroweak scale is chosen to set the dividing line between these scenarios. In particular, we find that inverse see-saw models of 
neutrino mass allow for light new physics interpretations of most precision leptonic observables, such as lepton universality, lepton flavor violation, but 
not for the electron EDM.
\end{abstract}

\section{Introduction}

Despite the success of the Standard Model (SM), there exists strong empirical evidence for new \textit{Beyond the Standard Model} (BSM) physics. This is required to explain  
neutrino mass, and dark matter, for example. 
In particular, since the first evidence of neutrino oscillations, the neutrino sector has been a driving force for BSM 
searches, where the relative smallness of the neutrino masses \cite{Abe:2015awa} compared to that of the charged leptons suggests a novel mass generation 
mechanism. 

	In broad terms, and given the lack of new physics discovered by the LHC thus far at the electroweak (EW) scale, it is useful to think in terms of two paradigms for understanding 
BSM physics. On the one hand we can imagine that new physics lies at energies above the SM energy scale. Typically these are best probed at the energy frontier at collider 
experiments. On the other hand, one can also imagine new physics being constituted by light degrees of freedom, which are necessarily weakly coupled to the SM. These types of 
theories are often best tested through precision observables, at the intensity frontier. Interestingly, 
neutrino masses can be understood equally well within those either paradigms. Indeed, a short-distance origin of light neutrino masses follows from the dimension-5 Weinberg  
operator $\alpha LLHH/\Lambda_{UV}$ \cite{Weinberg:1979sa}, which upon the Higgs developing a vev generates $m_\nu\sim\alpha \vev{H}^2/\Lambda_{UV}$. However, a long-distance 
origin of neutrino masses can also be understood at the renormalizable level, 
from the Yukawa coupling $yLHN_R$, with new light singlet states $N_R$, leading to Dirac neutrino masses, $m_\nu \sim y \vev{H}$, provided the coupling $y$ is sufficiently small \cite{Asaka:2005an}. Both regimes are of course connected, for example, by adding Majorana masses for $N_R$.


This contribution will review some recent work \cite{Dall:2015bba}, and explore the extent to which classes of low energy precision observables may be \textit{unambiguously} sensitive to \textit{either} high \textit{or} low energy physics. In other words, whether there are classes of observables that can uniquely \textit{point} to either UV or IR energy scales. We delineate four categories of {\it light} new physics theories as follows. 
\begin{enumerate}
\item The first category couples new neutral states via renormalizable operators. Such 
theories are exemplified by portal interactions, such as the Higgs or Neutrino portals \cite{Essig:2013lka}. 
\item The second category couples new hidden 
states  through anomaly free currents, such as $B-L$, and require no new charged degrees of freedom. 
\item The third category couples new states through anomaly free combinations, but which require non-zero SM charge assignments to satisfy the anomaly cancellation conditions. 
\item The last category includes new physics coupled through anomalous interactions. 
\end{enumerate}
We deem new physics as light if the masses are well below $m_W$, and consistent if they fall into 
the two first categories, so they do not require extra states above the EW scale. The third category requires additional charged scalars to generate masses for the hidden 
fields, and charged higgses below the EW scale are heavily constrained and pushed to the UV limit. The last category implies the introduction of a $UV$ cutoff, and the 
existence of UV scale states \cite{Preskill:1990fr}.

In the next section, a simple neutrino mass model is used to analyze various classes of leptonic precision observables, namely Lepton 
Flavor Violation (LFV), Lepton Universality (LU), and Lepton Electric Dipole Moments (LEDM). Our conclusion is that they are all sensitive to light 
physics, except the LEDMs which cannot be generated at the current sensitivity level within this class of models. A larger set of observables, including hadronic observables, has been considered in
\cite{Dall:2015bba}. A summary is given in Fig.~1.


\section{The Neutrino Model and Its Precision Constraints}

We consider a model of 3 left-handed active neutrinos $\nu_{l}$ with $l=e,\mu,\tau$, along with extra right-handed neutrinos $N_{Ri}$, and extra fermion singlets 
$N_{Sk}$,
\beq\label{eq: lagrangian}
	-\mathcal{L}_{\nu}\supset 
\left(\bery{ccc}\nu&N_R&N_S\ery\right)\left(\bery{ccc}0&m_{D}&\mu_{D}\\m_{D}&\epsilon_R&M_D\\\mu_{D}&M_D&\epsilon_S\ery\right)\left(\bery{c}\nu\\N_R\\N_S\ery\right).
\eeq
As a model of light new phyiscs, it falls in category 1 above, utilizing the neutrino portal, and requires no UV completion. 
In general, the mass entries are not diagonal and contain physical CP-phases, though we simplify the discussion by fine-tuning $m_D$ and $\mu_D$ such that they are nearly-diagonal 
and non-universal, i.e. $m_{D_{li}}\sim m_{D_l}\delta_{li}$, similarly for $\mu_D$. We assume the other matrices to be diagonal and 
universal, i.e. $M_{D_{ik}}\sim M_D\delta_{ik}$ and so on for $\epsilon_{R,S}$. We also assume the inverse 
see-saw regime \cite{Mohapatra:1986aw,GonzalezGarcia:1988rw}, where $\mu_{D}\sim\epsilon_R\sim0$, and $\epsilon_S\ll m_{D_l}\ll M_D$. 
At the lowest order in $m_{D_l}/M_D$, we can diagonalize the mass matrix in terms of a unitary transformation $U$ parametrized by the mixing angle $\Theta_l$ between hidden and 
visible states,

\beq
	\left(\bery{c}\nu_{l}\\N_R\\N_S\ery\right)\simeq\left(\bery{ccc}1&\Theta_l&i\Theta_l\\0&\frac{1}{\sqrt{2}}&-\frac{i}{\sqrt{2}}\\-\sqrt{2}
\Theta_l&\frac{1}{\sqrt{2}}&\frac{i}{\sqrt{2}}\ery\right)\left(\bery { c } \nu^m\\N_+\\N_-\ery\right),\quad\Theta_l=\frac{m_{D_l}}{\sqrt{2}M_{D}}.
\eeq
In the scenario of one $N_R$ and one $N_S$ for each active neutrino flavor, the associated mass spectrum consists of one light active neutrino $\nu^m$ and two heavier hidden 
neutrinos $N_\pm$ of masses $m_{\nu}\sim2\Theta_l^2\epsilon_S$, and $M_{\pm}\sim M_D\pm\epsilon_S/2$ respectively. For a model of light new physics, we take $m_{D_l},M_D\ll m_W$.

\paragraph{Lepton Flavor}
is weakly violated in the neutrino sector materialized in neutrino oscillations, but is otherwise an accidental symmetry in the charged lepton sector. There are various candidate 
channels to search for large sources of LFV \cite{Sawada:2013fba}. We will focus on $\mu\rightarrow e\gamma$ decay and $\mu-e$ conversion. The $\mu\rightarrow e\gamma$ 
decay is measured relative to the total muon decay rate 
\cite{Brooks:1999pu, Adam:2013mnn}, and is mediated at the one loop level \cite{Cheng:1976uq, Bjorken:1977br},
\beq\rm{Br}(\mu\rightarrow e\gamma)=\frac{3\alpha}{8\pi}\delta^2\approx 
\frac{3\alpha}{8\pi}\frac{M_D^4}{m_W^4}\Theta^2_{e}\Theta^2_{\mu}<5.7\cdot10^{-13},\quad\delta=\sum_{i=\nu,\pm}U_{ei}^*U_{\mu 
i}g\left(\frac{m_i^2}{m_W^2}\right).
\eeq

The $\mu-e$ conversion channel \cite{Weinberg:1959zz}, consists in the lepton flavor oscillation 
due to nuclear scattering, measured with respect to the muon capture rate $R_{\mu-e}=\Gamma(\mu-e)/\Gamma_{capture}$. This channel can be enhanced compared to 
the previous one thanks to the coherent nature of the low energy scattering. One contributing diagram \cite{Vergados:1985pq}, called \textit{photonic}, is the same as the on-shell 
decay except the photon line is connected to the nucleus. Another, the \textit{non-photonic} box diagram, can be competitive 
compared to the former due to the small quark masses running in the loop. In the light neutrino regime, the latter dominates, leading to
\beq
	R_{\mu-e}\approx\left(\frac{3G_Fm_W^2}{4\sqrt{2}\pi^2}\right)^2\frac{E_ep_e}{m_\mu^2}|F_{ch}|^2\rho_{coh}
\times\delta_\nu^2\approx26\frac { M_D^4 } { m_W^4 } \Theta^2_ { e } \Theta^2_ { \mu } \Leq7.0\cdot10^{-13},
\eeq
where $E_ep_e/m_\mu^2\sim1$ and the charge form factor $|F_{ch}|\sim0.5$. The limit comes from the SINDRUM II experiment using Gold \cite{Bertl:2006up}, for which the 
enhancement coherent factor $\rho_{coh}\sim1.6\cdot10^6$.

\paragraph{Lepton Universality}
tests flavour-independence of the coupling $g_{ll'}\overline l\gamma^\mu\nu_{l'} 
W_\mu$. In the SM with massless neutrinos, the coupling $g_{ll'}=g\delta_{ll'}$ is diagonal. When neutrinos acquire a mass, the flavor states $\nu_{l}$ become linear 
superpositions of mass eigenstates $\nu^m_i$, and the coupling senses all neutrino species through $g\sum_{i}U_{li}\overline l\gamma^\mu \nu^m_{i} W_\mu$. Among other variables, 
the 
ratio 
of leptonic tau decays $R_\tau=\Gamma(\tau\rightarrow\mu\nu\nu)/\Gamma(\tau\rightarrow e\nu\nu)$ tests LU, and is measured by BaBar to be \cite{Aubert:2009qj}
\beq
	R_\tau=\frac{\sum_{i,j}\Gamma(\tau\rightarrow\mu\nu^m_i\nu^m_j)}{\sum_{i,j}\Gamma(\tau\rightarrow e\nu^m_i\nu^m_j)}=0.9796\pm0.0052,\quad\Gamma(\tau\rightarrow 
e\nu^m_i\nu^m_j)=\frac{G_F^2m_\tau^5}{192\pi^2}|U_{\tau 
i}|^2|U_{lj}|^2I\left(\frac{m^2_l}{m^2_\tau},\frac{m^2_{\nu^m_{i,j}}}{m^2_\tau}\right),
\eeq
where $I(x,y)$ is a phase space kinematic function \cite{Abada:2013aba}. The deviation of the central value $0.9796$ from unity is due to 
radiative corrections. Ignoring those, we test to which extent the deviation from 1 is due to new physics, i.e. $\Delta R_\tau=1-\Gamma(\tau\rightarrow 
\mu\nu\nu)/\Gamma(\tau\rightarrow e\nu\nu)<0.0052$.

\btbl[t]
\label{fig: parameter plots}
\centering
\tabcolsep0pt
\btab{cc}
		\includegraphics[trim=0 50bp 0 0,clip,width = 7.2cm]{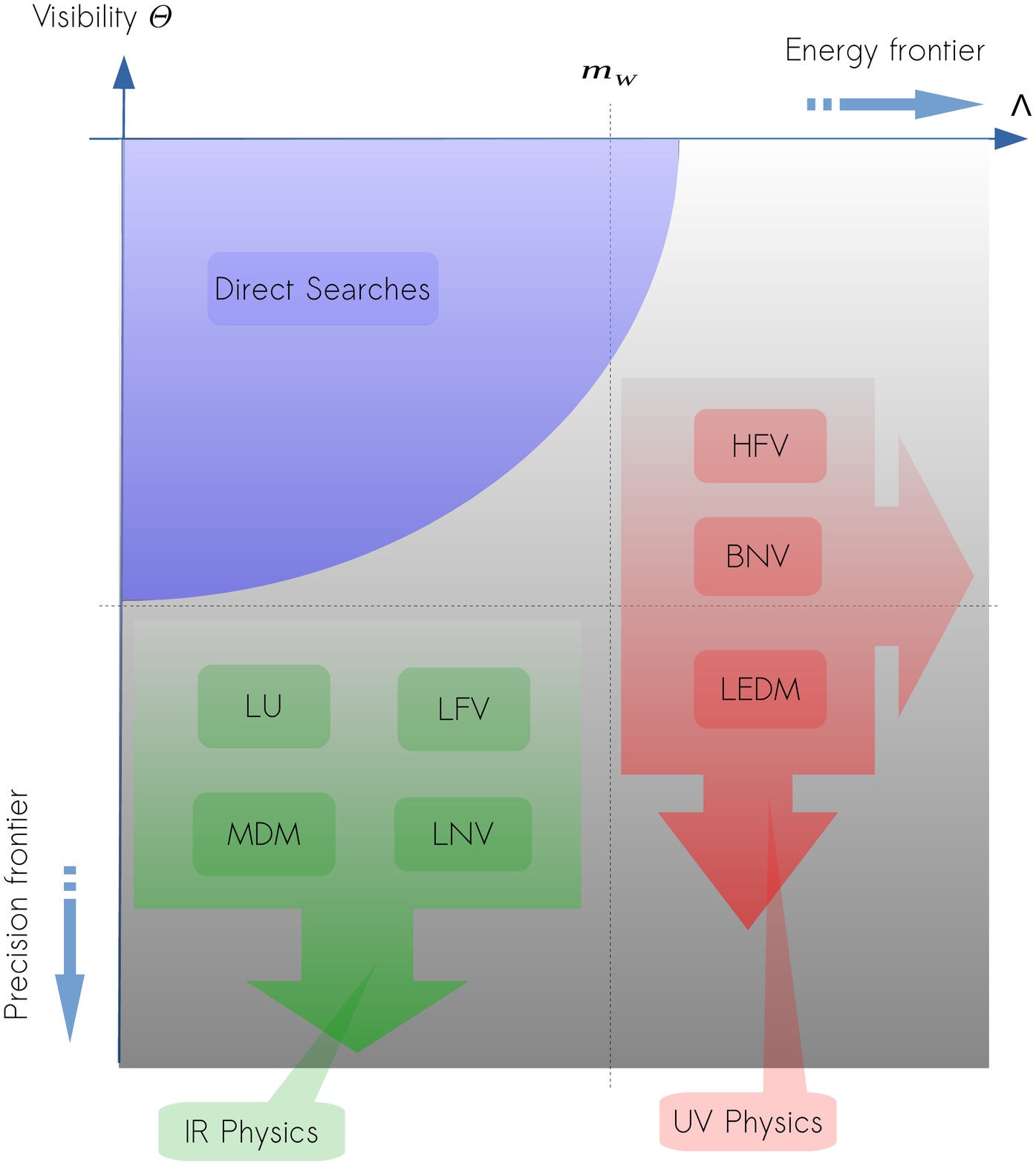}
		&
		\includegraphics[width = 7.5cm]{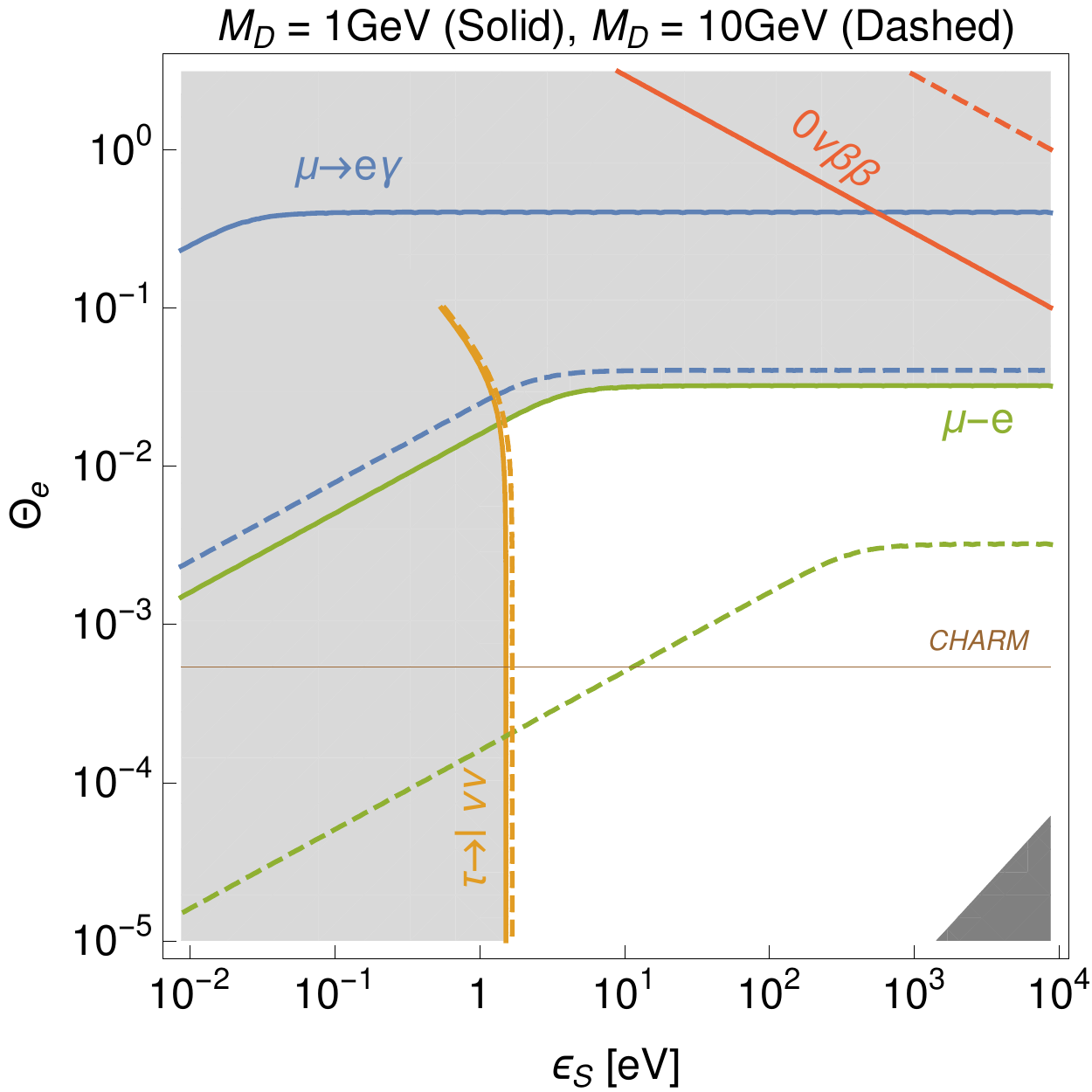}\\
		\multicolumn{2}{l}{\pbox{20cm}{\textbf{FIGURE 1.} Schematic sensitivity (left) of various precision observables, with the leptonic observables in green \\ being sensitive to light new physics as discussed here. The parameter space (right) associated with the neutrino \\ model in Eq.~(1), shows the current constraints, and viable regions.  
		The dark gray area marks the region\\ where our assumptions fail. The constraint from the CHARM experiment is marked in light brown  \cite{CHARM}.}}
		\\\hline
\etab
\etbl

\paragraph{Electric Dipole Moments}

are precision probes of CP-violation across a wide range of energy scales \cite{Pospelov:2005pr}.
Here, we focus on the electron EDM. In the conventional neutrino see-saw model, it can only arise at the 2 
W-loop level \cite{Ng:1995cs, Archambault:2004td, deGouvea:2005jj}, and depends on the neutrino mass square differences $\Delta m^2_{\nu}$ due to the GIM mechanism 
\cite{Glashow:1970gm}. As a 
result, the electron EDM generically scales as $\mathcal{O}(em_eG_F^2\Delta m^2_{\nu})\sim e\cdot \rm cm\cdot10^{-45}\Delta m^2_{\nu}/\rm eV^2$, which is orders of 
magnitude below the experimental upper limit $d_e < 8.7\times 10^{-29}e\cdot \rm cm$ \cite{Baron:2013eja}. It is possible to increase the estimate by using the full mass 
lagrangian Eq.~\eqref{eq: lagrangian} in a see-saw regime $\epsilon_{R,S}\gg m_D,\mu_D, M_D$, which leads to
\beq
 d_e\sim\left(5.8\cdot10^{-34}~ 
e\,\cdot\rm{cm}\right)\frac{m_{D}^2~\mu_{D}^2}{(\epsilon_R+\epsilon_S)^4}\frac{\epsilon_S^2-\epsilon_R^2}{\rm{GeV}^2}\sin(2\eta).
\eeq
In this regime, the active neutrino masses are controlled by the fine tuning of $m_D$ and $\mu_D$, $m_\nu\sim(m_D^2-\mu_D^2)/(\epsilon_R+\epsilon_S)$. 
Taking $\epsilon_S^2-\epsilon_R^2\simeq\epsilon_S^2\sim100~\rm GeV^2$, and $m_D/\epsilon_S,\mu_D/\epsilon_S\lesssim0.1$, we find the upper bound $d_e\lesssim10^{-36}~e\cdot\rm 
cm$, which is still well below 
the current experimental limit.

\section{Discussion}


The results of the above analysis are exhibited in Fig.~1, which shows that the inverse see-saw model provides a viable means of interpreting a range of precision measurements in  
terms of light new physics. We discussed LFV, LU, and LEDMs, and found that only lepton EDMs cannot be explained within the model, and would instead point to UV physics.

The study can be extended to hadron observables like hadron flavor violation, baryon number violation, 
hadron EDMs \cite{Dall:2015bba}. Since the hadronic sector does not allow for renormalizable couplings to neutral states, like the RHN in the lepton sector, there is no equivalent to the portal 
interactions, and a generic conclusion is that precision hadronic observables tend to test new short-distance physics.  
Note, however, that hadronic EDMs on the other hand seem to be \textit{ambiguous} pointers, due to the possibility of an explanation in terms of $\theta_{\rm QCD}$ which is a marginal coupling and can be generated anywhere above the QCD scale. 

\section{ACKNOWLEDGMENTS}

We are grateful to the Center for Theoretical Underground Physics and Related Areas (CETUP*) for their hospitality at the PPC2015 Conference. The work of  M.L. is supported in 
part by NSERC, Canada. The author would like to generously thank A. Ritz and M. Pospelov, for their collaboration and support for this work.

\bibliographystyle{aipnum-cp}
\bibliography{Proceeding_bib}

\end{document}